\documentclass{emulateapj}
\usepackage{epsfig}

\begin{document}

\title{Relativistic Hydrodynamic Flows Using Spatial and Temporal Adaptive
  Structured Mesh Refinement}

\author{Peng Wang, Tom Abel} 

\affil{Kavli Institute for Particle Astrophysics and Cosmology
  \\ Stanford Linear Accelerator Center and Stanford Physics
  Department, Menlo Park, CA 94025
  \\ Kavli Institute for Theoretical Physics, University of California, Santa Barbara, CA 93106} 
\author{Weiqun Zhang}
\affil{Kavli Institute for Particle Astrophysics and Cosmology
  \\ Stanford Linear Accelerator Center and Stanford Physics
  Department, Menlo Park, CA 94025} 
  \email{pengwang,
  tabel@stanford.edu, wqzhang@slac.stanford.edu}

\begin{abstract}
Astrophysical relativistic flow problems require high resolution
three-dimensional numerical simulations. In this paper, we describe a
new parallel three-dimensional code for simulations of special
relativistic hydrodynamics (SRHD) using both spatially and temporally
structured adaptive mesh refinement (AMR). We used the method of lines to
discretize the SRHD equations spatially and a total variation diminishing
(TVD) Runge-Kutta scheme for time integration. For spatial
reconstruction, we have implemented piecewise linear method (PLM),
piecewise parabolic method (PPM), third order convex essentially
non-oscillatory (CENO) and third and fifth order weighted essentially
non-oscillatory (WENO) schemes. Flux is computed using either direct
flux reconstruction or approximate Riemann solvers including HLL,
modified Marquina flux, local Lax-Friedrichs flux formulas and
HLLC. The AMR part of the code is built on top of the cosmological
Eulerian AMR code {\sl enzo}. We discuss the coupling
of the AMR framework with the relativistic solvers. Via various test problems, we emphasize the importance of resolution
studies in relativistic flow simulations because extremely high resolution is required especially when 
shear flows are present in the problem. We also present the results of two 3d simulations of astrophysical 
jets: AGN jets and GRB jets. Resolution study of those two cases further highlights the need of high resolutions
to calculate accurately relativistic flow problems.
\end{abstract}

\maketitle

\section{Introduction}

Relativistic flow problems are important in many astrophysical
phenomena including gamma-ray burst (GRB), active galactic nuclei
(AGN), as well as microquasar and pulsar wind nebulae, among
others. Apparent superluminal motion is observed in many jets of
extragalactic radio sources associated with AGN. According to the
currently accepted standard model, this implies the jet flow
velocities as large as $99\%$ of the speed of light
\citep{blandford77, begelman84}. Similar phenomena are also seen in
microquasars such as GRS 1915+105 \citep{mirabel94} and GRO J1655-40
\citep{tingay95} thus by similar arguments relativistic flows are
thought to play a role. In the case of GRB, the observed non-thermal
spectrum implies that the source must be optically thin, which can be
used to put a limit on the minimum Lorentz factor within those bursts
\citep{lithwick01}. This argument shows that the source of GRB must be
in highly relativistic motion. This conclusion is further confirmed by
the rapidly increasing number of GRB afterglow observations by the
Swift satellite \citep{gehrels04}. To understand physical processes in
those phenomena quantitatively, high resolution multi-dimensional
simulations are crucial.

Jim Wilson and collaborators pioneered the numerical solution of
relativistic hydrodynamics equations
\citep{wilson72,centrella84,hawley84a,hawley84b}. Starting with these
earliest papers this has typically been done in the context of general
relativistic problems such as accretion onto black holes and
supernovae explosions. The problem was recognized to be difficult to
solve when the Lorentz factor becomes large \citep{norman86} and a
solution with an implicit adaptive scheme was demonstrated in one
dimension. Unfortunately, this approach is not generalizable to
multi-dimensions. However, in the past two decades accurate solvers
based on Godunov's scheme have been designed that have adopted
shock-capturing schemes for Newtonian fluid to the relativistic fluid
equations in conservation form \citep[for a review
  see][]{martireview}. Such schemes, called high-resolution
shock-capturing (HRSC) methods, have been proven to be very useful in
capturing strong discontinuities with a few numerical zones without
serious numerical oscillations. We will discuss a number of them in
section 3.
 
Studies involving astrophysical fluid dynamics in general are
benefiting tremendously from using spatial and temporal adaptive
techniques. Smoothed particle hydrodynamics \citep{gingold77,lucy77}
is a classic example by being a Lagrangian method. Increasingly also
variants of \citet{berger89}'s structured adaptive mesh technique are being implemented. This is also true
in relativistic hydrodynamics \citep{hughes02, anninos05, zhang06,
  morsony06, meliani07} where certainly the work of \citet{hughes02}
showed that a serial AMR code could solve problems even highly
efficient parallel fixed grid codes would have difficulty with. In
this paper we discuss our implementation of
different hydrodynamics solvers with various reconstruction schemes as
well as different time integrators on top of the {\sl enzo} framework
previously developed for cosmology \citep{bryan97a, bryan97b, bryan01,
  o'shea04}. This new code we call {\sl r{\sl enzo}} is adaptive in
time and space and is a dynamically load balanced parallel using the
standard message passing interface. 


In the following we briefly summarize the equations being solved
before we give details on the different solvers we have
implemented. Section~4 discusses the adaptive mesh refinement strategy
and implementation. We then move on to describe various test problems
for relevant combinations of solvers, reconstruction schemes in one,
two and three dimensions, with and without AMR. Section~6 presents an
application of our code to three-dimensional relativistic and
supersonic jet propagation problem. Section~7 discusses 3d GRB jet simulation. 
We summarize our main conclusions in section~8.

\section{Equations of Special Relativistic Hydrodynamics}
\label{sec:equations}

The basic equations of special relativistic hydrodynamics (SRHD)
are conservation of rest mass and energy-momentum:
\begin{equation}
(\rho u^\mu)_{;\mu} = 0, \label{2.1}
\end{equation}
and
\begin{equation}
T^{\mu\nu}_{\ \ ;\nu} = 0, \label{2.2}
\end{equation}
where $\rho$ is the rest mass density measured in the fluid frame,
$u^{\mu} = W(1, v^i)$ is the fluid four-velocity (assuming the speed
of light $c=1$), $W$ is the Lorentz factor, $v^i$ is the coordinate
three-velocity, $T^{\mu\nu}$ is the energy-momentum tensor of the
fluid and semicolon denotes covariant derivative.

For a perfect fluid the energy-momentum tensor is
\begin{equation}
T^{\mu\nu} = \rho h u^{\mu} u^{\nu} + p g^{\mu\nu},
\end{equation}
where $h = 1 + \epsilon + p/\rho $ is the relativistic specific
enthalpy, $\epsilon$ is the specific internal energy, $p$ is the
pressure and $g^{\mu\nu}$ is the spacetime metric.

SRHD equations can be written in the form of conservation laws 
\begin{equation}
  \frac{\partial{U}}{\partial{t}} + \sum_{j=1}^{3}
  \frac{\partial{F^j}}{\partial{x^j}} = 0, \label{2.4}
\end{equation}
where the conserved variable $U$ is given by
\begin{equation}
  U = (D, S^1, S^2, S^3, \tau)^{T},
\end{equation} 
and the fluxes are given by
\begin{equation}
  F^j = (Dv^j, S^{1}v^{j}+p\delta^{j}_{\ 1},
  S^{2}v^{j}+p\delta^{j}_{\ 2}, S^{3}v^{j}+p\delta^{j}_{\ 3}, S^{j}-Dv^j)^{T}.
\end{equation}

\citet{anile89} has shown that system (\ref{2.4}) is hyperbolic for
causal EOS, i.e., those satisfying $c_s<1$ where the local sound speed
$c_s$ is defined as
\begin{equation}
c_s^2={1\over h}\left[{\partial p\over \partial \rho}+\left({p\over\rho^2}\right){\partial p\over \partial\epsilon} \right].\label{cs}
\end{equation}

The eigenvalues and left and right eigenvectors of the characteristic
matrix $\partial F/\partial U$, which are used in some of our
numerical schemes, are given by \citet{donat98}.

The conserved variables $U$ are related to the primitive variables by
\begin{eqnarray}
     D & = & \rho W, \label{d}\\
   S^j & = & \rho h W^2 v^j,  \label{sj}\\
  \tau & = & \rho h W^2 - p - \rho W, \label{tau}
\end{eqnarray}
where $j=1,2,3$.  The system (\ref{2.4})
are closed by an equation of state (EOS) given by $p =
p(\rho,\epsilon)$.  For an ideal gas, the EOS is,
\begin{equation}
  p = (\Gamma - 1) \rho \epsilon,
\end{equation} 
where $\Gamma$ is the adiabatic index. 

\section{Numerical Schemes for SRHD}
\label{sec:scheme}

\subsection{Time Integration}

We use method of lines to discretize the system (\ref{2.4}) spatially,
 \begin{eqnarray}
    {dU_{i,j,k}\over dt} &=& 
    - \frac{F^x_{i+1/2,j,k} -
      F^x_{i-1/2,j,k}}{\Delta x} \\ \nonumber
    &&- \frac{F^y_{i,j+1/2,k} -
      F^y_{i,j-1/2,k}}{\Delta y} \\ 
    &&- \frac{F^z_{i,j,k+1/2} - F^z_{i,j,k-1/2}}{\Delta z}\equiv L(U), 
    \label{ode}
\end{eqnarray}
where $i, j, k$ refers to the discrete cell index in $x, y, z$
directions, respectively. $F^{x}_{i \pm 1/2,j,k}$, $F^{y}_{i,j\pm
  1/2,k}$ and $F^{z}_{i,j,k\pm 1/2}$ are the fluxes at the cell
interface.

As discussed by \citet{shu88}, if using a high order scheme to
reconstruct flux spatially, one must also use the appropriate
multi-level total variation diminishing (TVD) Runge-Kutta schemes to
integrate the ODE system (\ref{ode}).  Thus we implemented the second
and third order TVD Runge-Kutta schemes coupled with AMR.

The second order TVD Runge-Kutta scheme reads,
\begin{eqnarray}
U^{(1)} & = & U^n + \Delta{t} L(U^n), 
  \\ 
        U^{n+1}&=&{1\over2}U^n+{1\over2}U^{(1)}+{1\over2}\Delta tL(U^{(1)}), \label{rk2}
\end{eqnarray}
and the third order TVD Runge-Kutta scheme reads,
\begin{eqnarray}
  U^{(1)} & = & U^n + \Delta{t} L(U^n) 
        \label{rk3-1} \\
  U^{(2)} & = & \frac{3}{4}U^n + \frac{1}{4}U^{(1)} + \frac{1}{4}
  \Delta{t} L(U^{(1)}) \label{rk3-2} \\ U^{n+1} & = & \frac{1}{3}U^n +
  \frac{2}{3}U^{(2)} + \frac{2}{3}\Delta{t} L(U^{(2)}) \label{rk3-3},
\end{eqnarray}
where
$U^{n+1}$ is the final value after advancing one time step
from $U^{n}$.

For an explicit time integration scheme, the time step is constrained
by the Courant-Friedrichs-Lewy (CFL) condition. The time step is
determined as
\begin{equation}
dt = C {\rm min}_i\left({\Delta x^i\over \alpha^i}\right), \label{dt}
\end{equation}
where $C$ is a parameter called CFL number and $\alpha^i$ is the local
largest speed of propagation of characteristics in the direction $i$
whose explicit expression can be found in \citet{donat98}.

\subsection{Reconstruction Method}

Generally speaking, there are two classes of spatially reconstruction
schemes (see e.g. LeVeque 2002). One is reconstructing the unknown
variables at the cell interfaces and then use exact or approximate
Riemann solver to compute the fluxes. Another is direct flux
reconstruction, in which we reconstruct the flux directly from
fluxes at the cell centers. To explore the coupling of different schemes
with AMR as well as exploring which method is most suitable for a
specific astrophysical problem, we implement several different
schemes in both classes.

To reconstruct unknown variables, we have implemented piecewise linear
method (PLM, Van Leer 1979), piecewise parabolic method (PPM, Colella
\& Woodward 1984, Mart\'i \& M\"uller 1996), the third-order convex
essentially non-oscillatory scheme (CENO, Liu \& Osher 1998). These are
used to reconstruct the primitive variables since reconstructing the
conserved variables can produce unphysical values in
SRHD. Furthermore, unphysical values of three-velocities may arise
during the reconstruction especially for ultrarelativistic flows. So
we either use $v^iW$ to do the reconstruction or we also reconstruct
the Lorentz factor and use it to renormalize the reconstructed
three-velocity when they are unphysical.

For direct flux reconstruction, we have implemented PLM and the third
and fifth order WENO scheme of \citet{jiang96}. Direct flux
reconstruction using WENO was first used to solve SRHD problems by
\citet{zhang06}. They showed that the fifth order WENO scheme works
well with the third order Runge-Kutta time integration. In our
implementation, we followed their description closely. For the PLM and CENO schemes, we used a generalized minmod slope
limiter \citep{kurganov00}. For given $v_{i-1}, v_i, v_{i+1}$,
$v_{i+1/2}=v_i+0.5{\rm minmod}(\theta(v_i-v_{i-1}),
0.5(v_{i+1}-v_{i-1}), \theta(v_{i+1}-v_i))$, where $1\le \theta \le
2$. For $\theta=2$ it reduces to the monotonized central-difference
limiter of \citet{van leer77}. We found that this generalized minmod
slope limiter behaves much better than a traditional minmod limiter
\citep{leveque02} especially for strong shear flows. In our
calculation, $\theta=1.5$ is used by default. For the PPM scheme, we
used the parameters proposed by \citet{marti96} for all the test
problems. For WENO, we used the parameters suggested in the original
paper of \citet{jiang96}.

\subsection{Riemann Solvers}

In the first class of reconstruction methods, given the reconstructed
left and right primitive variables at interfaces, the flux across each
interface is calculated by solving the Riemann problem defined by
those two states. An exact Riemann solver is quite expensive in SRHD
\citep{marti94, pons00}. Thus we have implemented several approximate
Riemann solvers including HLL \citep{harten83, schneider93,
  kurganov01}, HLLC \citep{toro94, mignone05a}, local Lax-Friedrichs
(LLF, Kurganov \& Tadmor 2000) and the modified Marquina flux
\citep{aloy99}.

The HLLC scheme is an extension of the HLL solver developed by
\citet{toro94} for Newtonian flow which was extended to two-dimensional
relativistic flows by \citet{mignone05a}. The improvement of HLLC over
HLL is restoring the full wave structure by constructing the two
approximate states inside the Riemann fan. The two states can be found
by the Rankine-Hugoniot conditions between those two states and the
reconstructed states. With this modification, HLLC indeed behaves
better than other Riemann solvers in some 1d (\S~5.1.7) and 2d
(\S~5.2) test problems. But when we apply HLLC to three dimensional
jet simulation, we found that HLLC suffers from the so called
``carbuncle" artifact well-known in the computational fluid dynamics
literature \citep{quirk94}.  We have used HLLC to run many other
two-dimensional test problems designed to detect the carbuncle
artifact and confirmed this shortcoming.  We found that the HLLC solver
is unsuitable for many multi-dimensional problems.  The discussion of
these problems will be presented elsewhere.  In this work, we will
only apply HLLC to two test problems showing that the HLLC solver has
less smearing at contact discontinuities than other schemes.

\Citet{lucas04} has compared the HLL scheme, the LLF scheme and the
modified Marquina flux formula using 1d and 2d test problems. They found those three schemes give
similar results for all their test problems. In our tests, we found
similar results. However, the modified Marquina flux formula is not as
stable as HLL in problems with strong transverse flows and LLF is more
diffusive than HLL. So in the following discussion we will only show
the results using HLL in most of the tests if there is no difference
among those three schemes.

In the following discussion, we will denote a specific hydro algorithm
by X-Y where X is the flux formula and Y is the reconstruction
scheme. For example, F-WENO5 denotes direct flux reconstruction using
fifth order WENO. We used the third order Runge-Kutta method for all
the tests in this work.

\subsection{Converting Conserved Variables to Primitive Variables}

Since primitive variables are needed in the reconstruction process,
after every RK time step, we need to convert conserved variables to
primitive variables. While conserved variables can be
computed directly from primitive variables using Eqs. (\ref{d}),
(\ref{sj}) and (\ref{tau}), the inverse operation is not
straightforward. One needs to solve a quartic equation for the ideal gas
EOS and a nonlinear equation for more complicated EOS. Iteration methods are used even for ideal gas EOS,
 because
computing the solution of a quartic is
expensive. Following \citet{aloy99}, we have used a Newton-Raphson
(NR) iteration to solve a nonlinear equation for pressure to recover
primitive variables from conserved variables. Typically, the NR
iteration needs only $2$ to $3$ steps to converge.

\subsection{Curvilinear coordinates}

We have also implemented cylindrical and spherical coordinates following the
description of Zhang \& MacFadyen (2006). This affects three parts of
the code.  Firstly, the geometric factors are incorporated into the
flux when updating the conserved variables. Secondly, there will be
geometric source terms. Thirdly, the flux correction in AMR (\S~4.4)
is modified by geometric factors.

\section{Adaptive Mesh Refinement}
\label{sec:amr}

\subsection{Overview}

Structured adaptive mesh refinement (AMR) was developed by \citet{berger84} and
\citet{berger89} to achieve high spatial and temporal resolution in
regions where fixed grid resolution is insufficient. In structured AMR, a subgrid will be created in regions of its
parent grid needing higher resolution. The hierarchy of grids is
a tree structure. Each grid is evolved as a separate initial boundary
value problem, while the whole grid hierarchy is evolved recursively.

{\sl r{\sl enzo}} is built on top of the AMR framework of {\sl enzo}
\citep{bryan97a, o'shea04}. {\sl enzo}'s implementation of AMR follows
closely the Berger \& Colella paper and has been shown to be very
efficient for very high dynamic range cosmological simulations (see
e.g. Abel et al. 2002). The pseudocode of the main loop for the second order
Runge-Kutta method reads,
\begin{eqnarray}
&&{\rm EvolveLevel} \ \ l  \nonumber \\ \nonumber
&&\ \ \ \ {\rm SetBoundaryCondition} \\ \nonumber
&& \ \ \ \ {\rm while}(t_l < t_{l-1}) \\ \nonumber
&& \ \ \ \ \ \ \ \ {\rm ComputeTimeStep} \ \ dt_l \\ \nonumber
&& \ \ \ \ \ \ \ \ {\rm for \ \ every \ \ grid \ \ patch \ \ on \ \ this \ \ level} \\ \nonumber 
&& \ \ \ \ \ \ \ \ \ \ \ \  {\rm Runge-Kutta  \ \ first \ \ step:} \ \ {\rm Eq}. (14) \\ \nonumber
&& \ \ \ \ \ \ \ \ \ \ \ \ \ \ \ \ {\rm ComputeFlux} \\ \nonumber
&& \ \ \ \ \ \ \ \ \ \ \ \ \ \ \ \ \ \ \ \  {\rm SweepX, Y, Z} \\ \nonumber
&& \ \ \ \ \ \ \ \ \ \ \ \ \ \ \ \ \ \ \ \ \ \ \ \  {\rm ChooseHydroAlgorithm} \\ \nonumber
&& \ \ \ \ \ \ \ \ \ \ \ \ \ \ \ \ \ \ \ \ \ \ \ \  {\rm SaveSubgridFlux} \\ \nonumber 
&& \ \ \ \ \ \ \ \ \ \ \ \ \ \ \ \  {\rm UpdateConservedVariables} \\ \nonumber
&& \ \ \ \ \ \ \ \ \ \ \ \ \ \ \ \  {\rm ConservedToPrimitive} \\ \nonumber
&& \ \ \ \ \ \ \ \ \ \ \ \ {\rm UpdateTime:} \ \ t_l = t_l + dt_l \\ \nonumber
&& \ \ \ \ \ \ \ \ {\rm SetBoundaryCondition} (\S~4.2) \\ \nonumber
&& \ \ \ \ \ \ \ \ {\rm for \ \ every \ \ grid \ \ patch \ \ on \ \ this \ \ level} \\ \nonumber 
&& \ \ \ \ \ \ \ \ \ \ \ \  {\rm Runge-Kutta \ \ second \ \ step:} \ \ {\rm Eq}. (15) \\ \nonumber
&& \ \ \ \ \ \ \ \ \ \ \ \ \ \ \ \  {\rm ComputeFlux} \\ \nonumber
&& \ \ \ \ \ \ \ \ \ \ \ \ \ \ \ \ {\rm UpdateConservedVariables} \\ \nonumber
&& \ \ \ \ \ \ \ \ \ \ \ \ \ \ \ \ {\rm ConservedToPrimitive} \\ \nonumber
&& \ \ \ \ \ \ \ \ {\rm SetBoundaryCondition} \\ \nonumber
&& \ \ \ \ \ \ \ \ {\rm EvolveLevel} \ \ l+1 \\ \nonumber
&& \ \ \ \ \ \ \ \ {\rm UpdateFromFinerGrid} \ \ (\S~4.4) \\ \nonumber
&& \ \ \ \ \ \ \ \ {\rm FluxCorrection} \ \ (\S~4.4) \\ \nonumber
&& \ \ \ \ \ \ \ \ {\rm RebuildHierarchy \ \ for \ \ level} \ \ l \\ \nonumber
 \end{eqnarray}
 
The RebuildHierarchy function called at the end of every time step is
at the heart of AMR. Its
pseudocode as implemented originally in {\sl enzo} reads,
\begin{eqnarray}
&&{\rm RebuildHierarchy \ \ for \ \ level } \ \ l \\ \nonumber
&& \ \ \ \ {\rm for \ \ i_{level}=} \ \ l \ \ {\rm to \ \ MaximumLevel}-1 \\ \nonumber
&& \ \ \ \ \ \ \ \ {\rm for \ \ every \ \ grid \ \ on \ \ i_{level}} \\ \nonumber
&& \ \ \ \ \ \ \ \ \ \ \ \ {\rm FlagCellsForRefinement} \ \ (\S~4.3) \\ \nonumber
&& \ \ \ \ \ \ \ \ \ \ \ \ {\rm CreateSubgrids} \\ \nonumber
&& \ \ \ \ \ \ \ \ \ \ \ \ {\rm AddLevel(i_{level}+1)} \\ \nonumber
&& \ \ \ \ \ \ \ \ {\rm for \ \ every \ \ new \ \ subgrid}\\ \nonumber
&& \ \ \ \ \ \ \ \ \ \ \ \ {\rm InterpolateFieldValuesFromParent} \ \ (\S~4.2) \\ \nonumber
&& \ \ \ \ \ \ \ \ \ \ \ \ {\rm CopyFromOldSubgrids} \\ \nonumber
&& \ \ \ \ \ \ \ \ {\rm LoadBalanceGrids} \\ \nonumber
\end{eqnarray}

\subsection{Interpolation}

When a new subgrid is created, the initial values on that grid are
obtained by interpolating spatially from its parent grid. In this
case, we apply the conservative second order order interpolation
routine provided by {\sl enzo} to conserved variables.  But in this
process sometimes the interpolated values can violate the constraint
$(\tau+D)^2>S^2+D^2$. If this happens, we will then use first order
method for that subgrid.

Before the first Runge-Kutta step for a grid at level $l$ (in the
following discussion, we use the convention that top grid has level
$0$), we will need the boundary condition at time $t_{l}$, which is derived
 by interpolating from its parent grid. Then at the later steps
of Runge-Kutta scheme, one needs the boundary condition at time
$t_l+dt_l$. Since the variables of its parent grid has already been
evolved to time $t_{l-1}+dt_{l-1}$, which is greater than time
$t_l+dt_l$, we can obtain the boundary conditions at time $t_l+dt_l$
for a grid at level $l$ by interpolating both temporally and spatially
from its parent grid. There are two exceptions to this
procedure. First, if a cell of fine grid abuts the box boundary, then
we just use the specified boundary condition for that cell. Second, if
a cell abuts another grid at the same level, we copy the
value from that grid. Because of the above mentioned problem for
interpolating conserved variables, when interpolating boundary values,
we apply the second order interpolation to primitive variables. Since
for ultrarelativistic flows spatially interpolating three-velocity can
lead to unphysical values, we also interpolate the Lorentz factor
and then use it to renormalize the interpolated three-velocity.

\subsection{Refinement Criteria}

In the test problems discussed in section \ref{sec:test}, we mainly
used two general purpose refinement criteria that have been widely
used in AMR code \citep{zhang06}.

In the first one, we compute the slope 
\begin{equation}
S_i = \frac{|u_{i+1}-u_{i-1}|}{{\rm max}(|u_i|, \epsilon)} ,
\end{equation}
where $u_i$ is typically density, pressure and velocities, $\epsilon$
is a small number typically taken to be $10^{-10}$. When $S_i$ is
larger than a minimum slope, typically $1$, a cell will be flagged for
refinement.

In the second one, for every cell we compute
\begin{equation}
  E_i = \frac{|u_{i+2} - 2 u_i + u_{i-2}|}{|u_{i+2}-u_i| +
    |u_i-u_{i-2}| + \epsilon (|u_{i+2}| + 2|u_i| + |u_{i-2}|)},
\end{equation}
which is the ratio of the second and first derivatives with a safety
factor in the denominator. Unless otherwise stated, we use $\epsilon =
0.01$. When $E_i$ is larger than a critical value $E_{ref}$, a cell
will be flagged for refinement. Typically we use $E_{ref}=0.8$.

To fully exploit AMR, it is desirable to design more
specific refinement criterions that are most efficient for a specific
astrophysical problems. 

\subsection{Flux Correction}

When a cell is overlayed by a finer level grid, then the coarse grid
value is just the conservative average of the fine grid values. On the
other hand, when a cell abuts a fine grid interface but is not itself
covered by any fine grid, we will do flux correction for that cell,
i.e. we will use the fine grid flux to replace the coarser grid flux
in the interface abutting the fine grid.(see Berger \& Colella 1989
for more detailed description of flux correction). For this purpose,
note that the second order Runge-Kutta method can be rewritten as
\begin{equation}
U^{n+1}=U^n+{1\over2}\Delta tL(U^n)+{1\over2}\Delta tL(U^{(1)}),
\end{equation}
and the third order Runge-Kutta method can be rewritten as
\begin{equation}
U^{n+1}=U^n+{1\over6}\Delta tL(U^n)+{1\over6}\Delta tL(U^{(1)})+{2\over3}\Delta tL(U^{(2)}).
\end{equation}
Thus for example, when we do flux correction in the $x$-direction for
interface $i+1/2$, we will use
$F^x_{i+1/2}(U^n)/2+F^x_{i+1/2}(U^{(1)})/2$ and
$F^x_{i+1/2}(U^n)/6+F^x_{i+1/2}(U^{(1)})/6+2F^x_{i+1/2}(U^{(2)})/3$ to
correct the coarser grid conserved variables for the second and third
order Runge-Kutta method, respectively.

\subsection{Parallelism}

{\sl r{\sl enzo}} uses the {\sl enzo} parallel framework which uses
dynamically load balancing using the Message Passing Interface (MPI)
library. At run time, the code will move grids among processors
according to the current load of every processor to achieve a balanced
distribution of computational load among processors. The computational
load of a processor is defined as the total number of active cells on
that processor and level.

\section{Code Tests}
\label{sec:test}


\begin{figure}
  \epsscale{1.2}
  \plotone{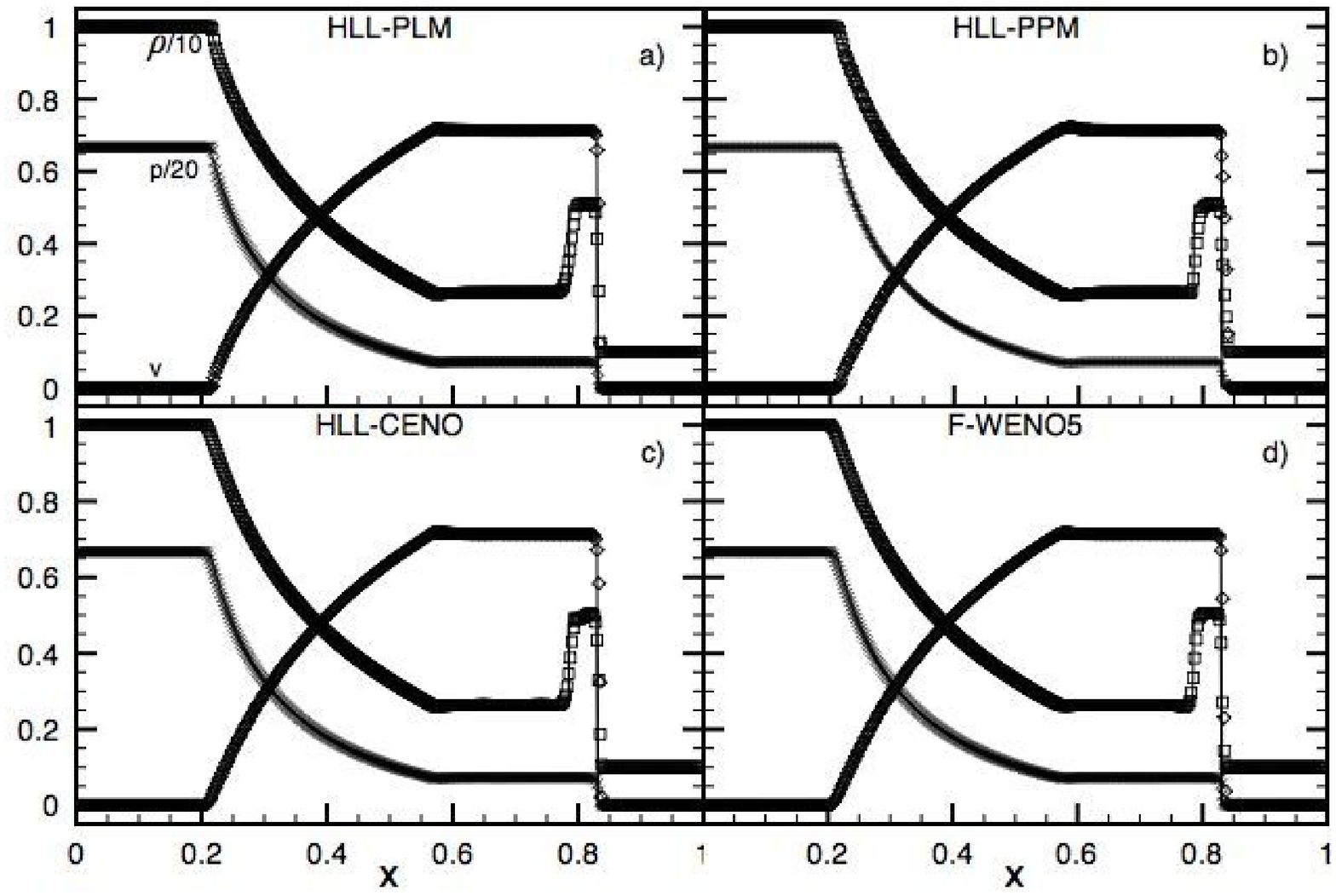}
    \caption{Relativistic blast wave I at $t=0.4$ with uniform
      resolution $N=400$ for (a) HLL-PLM, (b) HLL-PPM, (c) HLL-CENO
      and (d) F-WENO5. Numerical profiles of density (squares),
      pressure (cross signs) and velocity (diamonds) are shown as well
      as the analytical solution (solid lines). The CFL number used is
      $0.5$.}\label{fig:riemann1}
\end{figure}

\begin{figure}
 \epsscale{1.2}
 \plotone{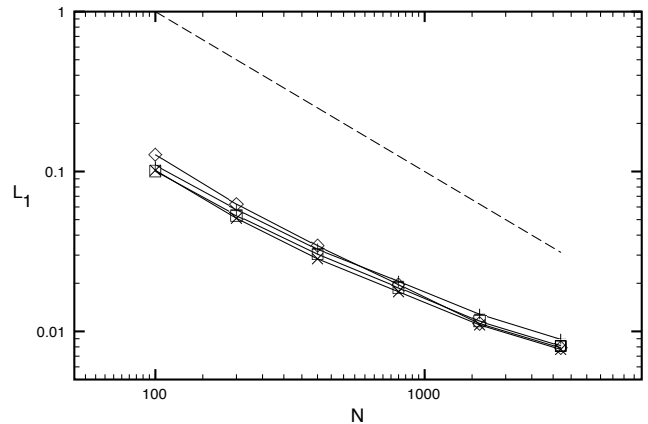}
    \caption{L1 errors in rest mass density for the relativistic blast
      wave I problem for six different uniform grid resolutions. The
      symbols denote HLL-PLM (plus signs), HLL-PPM (diamonds),
      HLL-CENO (squares) and F-WENO5 (cross signs). The dashed line
      indicates first order of global
      convergence.}\label{fig:l1_riemann1}
\end{figure}

\subsection{One Dimensional Test}
Relativistic Riemann problems have analytical solutions
\citep{pons00}, thus they are ideal for tesing SRHD codes. In the
following discussion, subscript $L$ and $R$ refer to the left and
right initial states, respectively. The initial discontinuity is
always at $x=0.5$. We will report the error between numerical
solutions and analytical solutions using $L_1$ norm defined as $L_1 =
\Sigma_i|u_i-u(x^i)|\Delta x^i$ where $u_i$ is the numerical solution,
$u(x^i)$ is the analytical solution and $\Delta x^i$ is the cell
width.

\subsubsection{Relativistic Blast Wave I}
This test and the following one are fairly standard and all modern
SRHD codes can match the analytical solution quite well (see Mart\'t
\& M\"uller 2003 for a summary of different codes' performance on
those two tests).

The initial left and right states for this problem are $p_L = 13.33,
\rho_L = 10.0, v_L=0.0$ and $p_R = 10^{-6}$, $\rho_R=1.0$, $v_R =
0.0$. The gas is assumed to be ideal with an adiabatic index $\Gamma =
5/3$. The initial discontinuity gives rise to a transonic rarefaction
wave propagating left, a shock wave propagating right and a contact
discontinuity in between. This problem is only mildly relativistic
with a post-shock velocity $0.72$ and shock velocity 0.83. The results
using four hydro solvers are shown in Fig. \ref{fig:riemann1}. The CFL
number used is $0.5$. The $L_1$ errors are shown in
Fig. \ref{fig:l1_riemann1}, from which we can see the four schemes
behave essentially identical for this problem. The order of global
convergence rate is about 1 for all four schemes, which is consistent
with the fact that there are discontinuities in the problem.

\begin{figure}
  \epsscale{1.2}
  \plotone{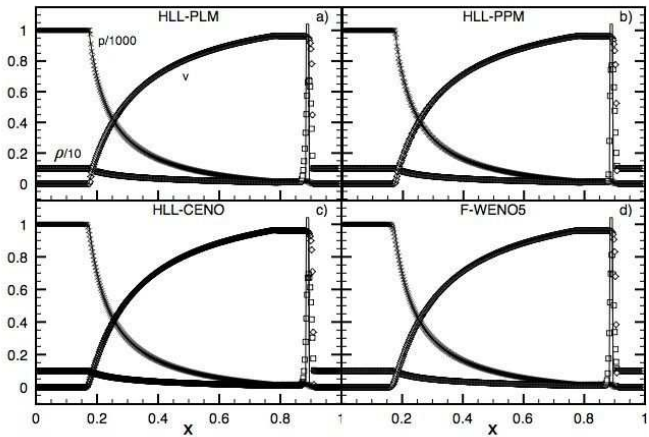}
    \caption{Relativistic blast wave II at $t=0.4$ with uniform
      resolution $N=400$ for (a) HLL-PLM, (b) HLL-PPM, (c) HLL-CENO
      and (d) F-WENO5. Numerical profiles of density (squares),
      pressure (cross signs) and velocity (diamonds) are shown as well
      as the analytical solution (solid lines). The CFL number used is
      $0.5$.}\label{fig:riemann2}
\end{figure}

\begin{figure}
  \epsscale{1.2}
  \plotone{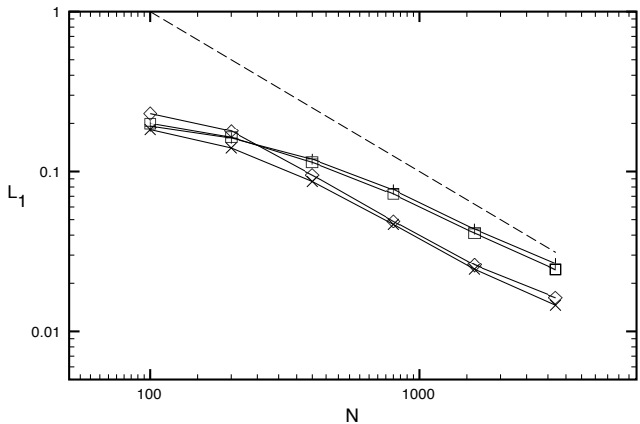}
    \caption{L1 errors in rest mass density for the relativistic blast
      wave II problem for six different uniform grid resolutions. The
      symbols denote HLL-PLM (plus signs), HLL-PPM (diamonds),
      HLL-CENO (squares) and F-WENO5 (cross signs). The dashed line
      indicates first order of global
      convergence.}\label{fig:l1_riemann2}
\end{figure}

\subsubsection{Relativistic Blast Wave II}
The initial left and right states for this problem are $p_L = 1000.0,
\rho_L = 1.0, v_L=0.0$ and $p_R = 10^{-2}$, $\rho_R = 1.0$, $v_R =
0.0$. The gas is assumed to be ideal with an adiabatic index $\Gamma =
5/3$. This test is more relativistic than the previous one. While the
wave structure is the same, the thermodynamically relativistic initial
left state gives rise to a relativistic shock propagating at a Lorentz
factor $W\approx 6$ and a very thin dense shell behind the shock with
width $\approx 0.01056$ at $t=0.4$. The CFL number used is $0.5$. The
results using four hydro algorithms are shown in
Fig. \ref{fig:riemann2}.  The $L_1$ errors are shown in
Fig. \ref{fig:l1_riemann2}. We can see that for this problem PPM and
WENO have smaller $L_1$ errors than PLM and CENO. This is due to their ability to
better resolve the thin shell.

\begin{figure}
  \epsscale{1.2}
  \plotone{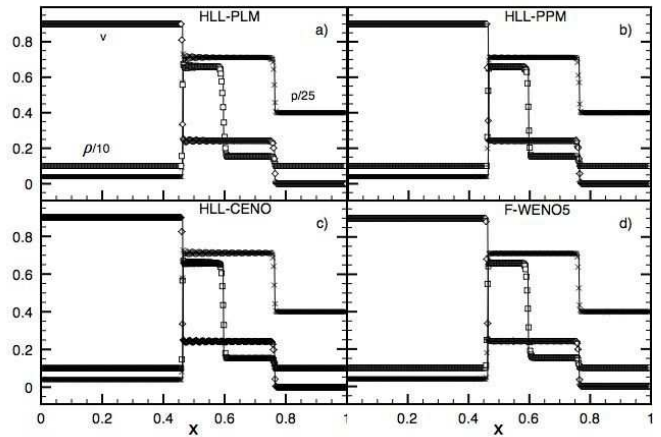}
    \caption{Planar jet problem at $t=0.4$ with uniform resolution
      $N=400$ for (a) HLL-PLM, (b) HLL-PPM, (c) HLL-CENO and (d)
      F-WENO5. Numerical profiles of density (squares), pressure
      (cross signs) and velocity (diamonds) are shown as well as the
      analytical solution (solid lines). The CFL number used is
      $0.5$.}\label{fig:riemann3}
\end{figure}

\begin{figure}
  \epsscale{1.2}
  \plotone{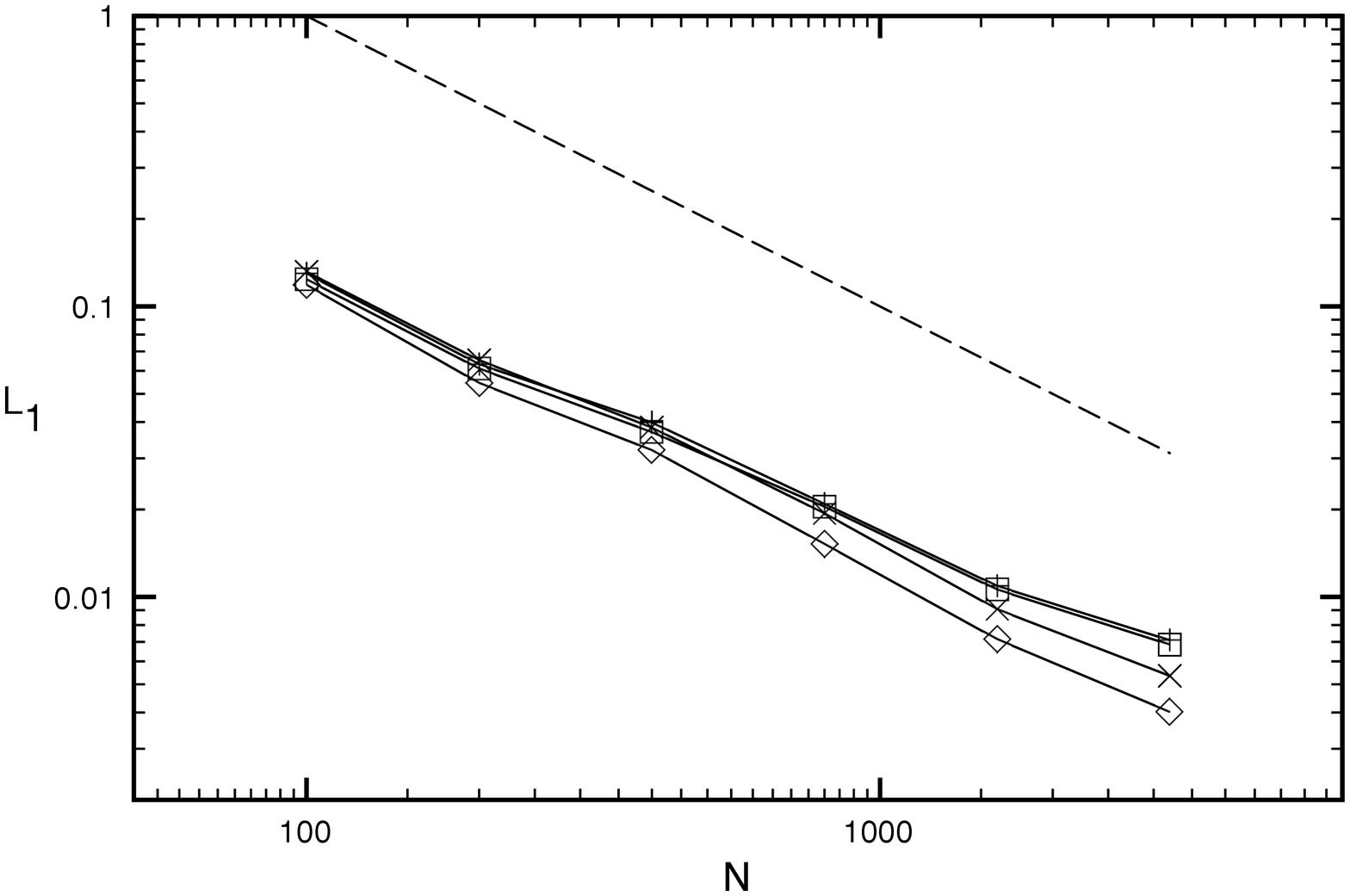}
    \caption{L1 errors in rest mass density for the planar jet problem
      for six different uniform grid resolutions. The symbols denote
      HLL-PLM (plus signs), HLL-PPM (diamonds), HLL-CENO (squares) and
      F-WENO5 (cross signs). The dashed line indicates first order of
      global convergence.}\label{fig:l1_riemann3}
\end{figure}

\begin{figure}
\epsscale{1.2} 
\plotone{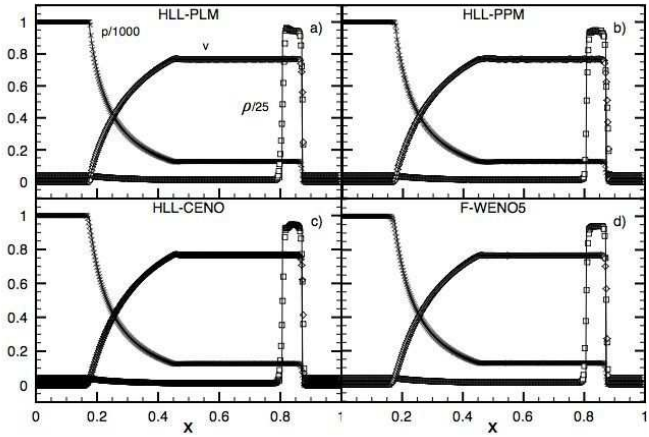}
  \caption{Blast wave with transverse velocity problem I at $t=0.4$
      with uniform resolution $N=400$ for (a) HLL-PLM, (b) HLL-PPM,
      (c) HLL-CENO and (d) F-WENO5. Numerical profiles of density
      (squares), pressure (cross signs) and velocity (diamonds) are
      shown as well as the analytical solution (solid lines). The CFL
      number used is $0.5$.}\label{fig:transverse_c}
\end{figure}

\subsubsection{Planar Jet Propagation}
The initial left and right states for this test are $p_L = 1.0, \rho_L
= 1.0, v_L=0.9$ and $p_R = 10.0$, $\rho_R =1.0$, $v_R = 0.0$. The gas
is assumed to be ideal with an adiabatic index $\Gamma = 4/3$. This
test mimics the interaction of a planar jet head with the ambient
medium. The decay of the initial discontinuity gives rise to a strong
reverse shock propagating to the left, a forward shock propagating to
the right and a contact discontinuity in between. The results are
shown in Fig. \ref{fig:riemann3}. The CFL number is $0.5$. The $L_1$
errors are shown in Fig. \ref{fig:l1_riemann3}.  As can be seen in
Figs. \ref{fig:riemann3} \& \ref{fig:l1_riemann3}, for this
problem PPM and WENO behave better than PLM and CENO: there is almost
no oscillation behind the reverse shock and they capture the contact
discontinuity with fewer cells. Especially PPM captures the contact
discontinuity with only $4$ cells and has the smallest $L_1$ error.

\subsubsection{Blast Wave with Transverse Velocity I}
For this and the following two problems, we will consider non-zero
transverse velocities in the initial states. The initial state is
identical to blast wave problem II except the presence of transverse
velocities. Those problems were first discussed analytically by
\citet{pons00}. Since then various groups have shown that when
transverse velocities are non-zero, in some cases those problems
become very difficult to solve numerically unless very high spatial
resolution is used \citep{mignone05b, zhang06, morsony06}. In
realistic astrophysical phenomena transverse velocities are usually
very important (see e.g. Aloy \& Rezzolla 2006), thus solving those
problems accurately is of great importance.

\begin{figure}
\epsscale{1.2}
  \plotone{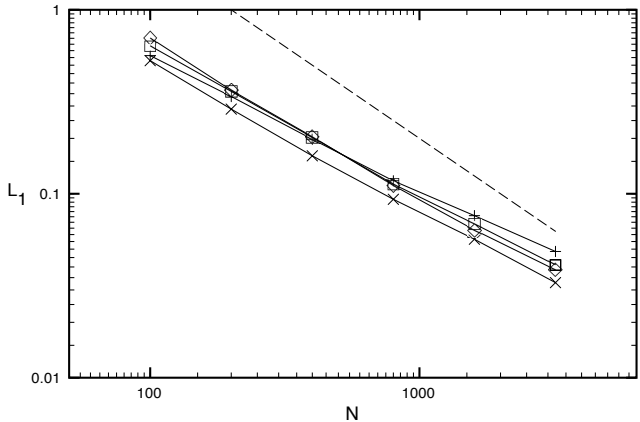}
    \caption{L1 errors in rest mass density for the blast wave with
      transverse velocity problem I for six different uniform grid
      resolutions. The symbols denote HLL-PLM (plus signs), HLL-PPM
      (diamonds), HLL-CENO (squares) and F-WENO5 (cross signs). The
      dashed line indicates first order of global
      convergence.}\label{fig:l1_bt1}
\end{figure}

As an easy first case, we will consider non-zero transverse velocity
only in the low pressure region. The initial left and right states are
$p_L = 1000.0, \rho_L = 1.0, v_{xL}=0.0, v_{yL} = 0.0$ and $p_R =
10^{-2}, \rho_R = 1.0, v_{xR} = 0.0, v_{yR} = 0.99$. The gas is
assumed to be ideal with an adiabatic index $\Gamma = 5/3$. The
results are shown in Fig. \ref{fig:transverse_c}. The CFL number is
$0.5$. The $L_1$ errors are shown in
Fig. \ref{fig:l1_bt1}. We can see that all four hydro
algorithms behaves similarly well, except that PLM and CENO
shows some small oscillations around the contact discontinuity.

\begin{figure}
  \epsscale{1.3}
  \plotone{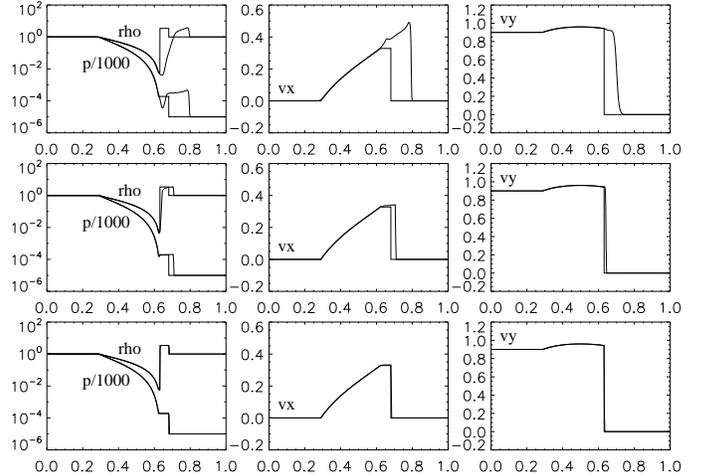}
    \caption{Blast wave with transverse velocity problem II at $t=0.4$
      with uniform resolution $N=400$ (top), 4 refinement levels with
      a refinement factor of 2 (middle, equivalent resolution 6400) and 3
      refinement levels with a refinement factor of 4 (bottom, equivalent
      resolution 25600). The hydro solver used is HLL-PLM. Thick
      lines are numerical solution while thin lines are analytical
      solutions. The CFL number used is
      $0.4$}\label{fig:transverse_d}
\end{figure}

\subsubsection{Blast Wave with Transverse Velocity II}
Next, we consider non-zero transverse velocity in the high
pressure region. In this case, the problem becomes more difficult to solve
numerically \citep{mignone05b, zhang06}. The initial left and right
states for this problem are $p_L = 1000.0, \rho_L = 1.0, v_{xL}=0.0,
v_{yL} = 0.9$ and $p_R = 10^{-2}, \rho_R = 1.0, v_{xR} = 0.0, v_{yR} =
0.0$. The gas is assumed to be ideal with an adiabatic index $\Gamma =
5/3$.  The high pressure region is connected to the intermediate state
by a rarefaction wave. Since the initial normal velocity in the high
pressure region is zero, the slope of the adiabat increases rapidly
with transverse velocity, thus a large initial transverse velocity
will lead to a small intermediate pressure and a small mass flux.

\begin{figure}
  \epsscale{1.2}
  \plotone{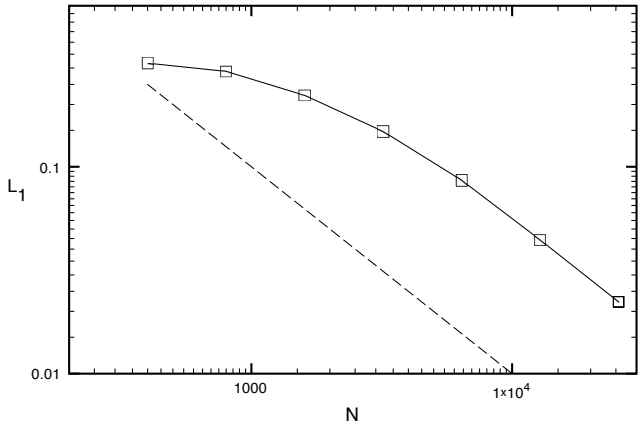}
    \caption{L1 errors in rest mass density for the blast wave with
      transverse velocity problem II for seven different equivalent
      grid resolutions using AMR. The solid lines with square signs
      are for HLL-PLM. The dashed line indicates first order of global
      convergence.}\label{fig:l1_transverse_d}
\end{figure}

The results using a uniform grid and two AMR runs are shown in
Fig. \ref{fig:transverse_d}. The hydro solver used for this figure is
HLL-PLM. The CFL number is $0.4$.  The $L_1$ error are shown in
Fig. \ref{fig:l1_transverse_d}. It can be seen that for the run with
400 uniform grid cells, the numerical solution is inadequate, as
previously found by \citet{mignone05b}. This is mainly due to the poor
capture of the contact discontinuity. We have tried to run this
problem with various algorithms but only obtained accurate solutions
by dramatically increasing the resolution.

\begin{figure}
  \epsscale{1.3}
  \plotone{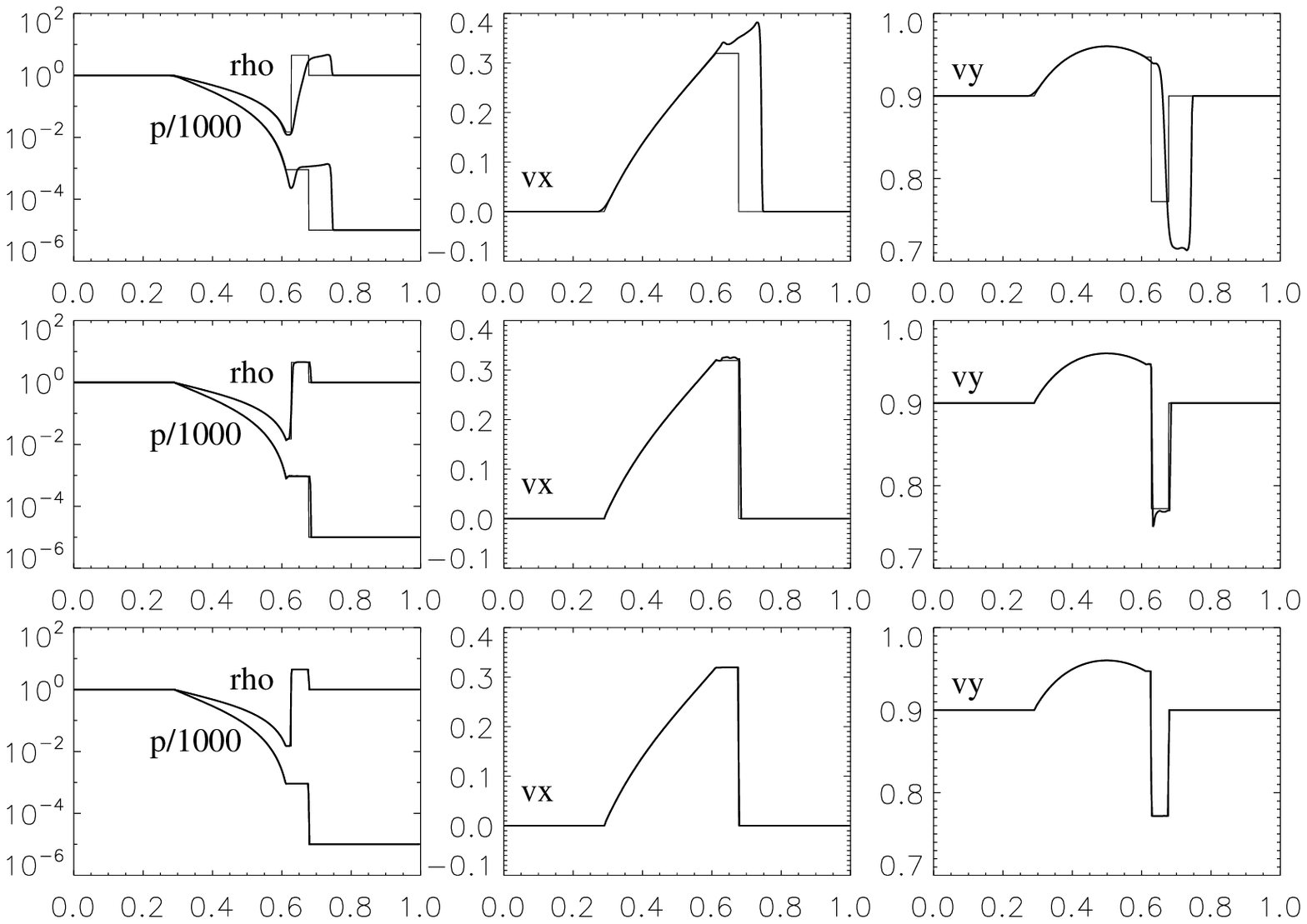}
    \caption{Blast wave with transverse velocity problem III at
      $t=0.4$ with uniform resolution $N=400$ (top), 4 refinement
      levels with a refinement factor of 2 (middle, equivalent resolution
      6400) and 4 refinement levels with a refinement factor of 3 (bottom,
      equivalent resolution 25600). The hydro solver used is
      HLL-PLM. Thick lines are numerical solutions while thin lines
      are analytical solutions. The
      CFL number used is $0.5$}\label{fig:transverse_e}
\end{figure}

\begin{deluxetable}{cccc}
\tablecaption{Equivalent resolution and the actual number of cells used in
  Blast Wave problems with Transverse Velocity II (BT II) and III (BT
  III).
\label{tab:1}}
\tablewidth{0pt}
\tablehead{
\colhead{Equivalent Resolution}    &   \colhead{BT II}  &  
\colhead{BT III} 
}
\startdata
400 & 400 & 400 \\
800 & 448 & 421 \\
1600 & 455 & 442 \\
3200 & 470 & 468 \\
6400 & 501 & 473 \\
12800 & 518 & 476 \\
25600 & 562 & 520
\enddata

\end{deluxetable}

\subsubsection{Blast Wave with Transverse Velocity III}
Now we introduce transverse velocity in both region. The initial left
and right states for this problem are $p_L = 1000.0, \rho_L = 1.0,
v_{xL}=0.0, v_{yL} = 0.9$ and $p_R = 10^{-2}, \rho_R = 1.0, v_{xR} =
0.0, v_{yR} = 0.9$. The gas is assumed to be ideal with an adiabatic
index $\Gamma = 5/3$.  This problem is more difficult than the
previous one due to the formation of an extremely thin shell
between the rarefaction wave tail and the contact discontinuity
\citep{zhang06}.

\begin{figure}
  \epsscale{1.2}
  \plotone{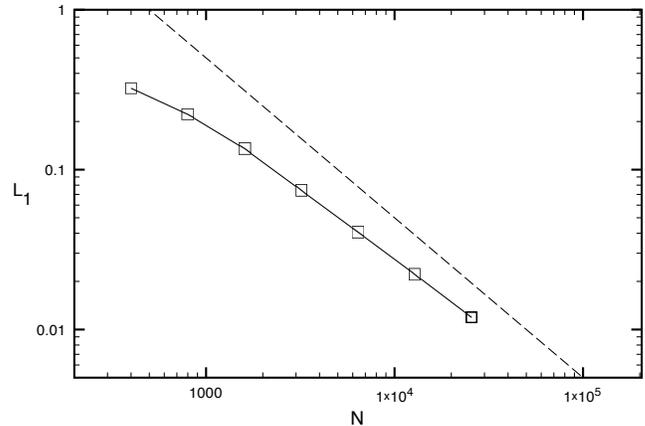}
    \caption{L1 errors in rest mass density for the blast wave with
      transverse velocity problem III for seven different equivalent
      grid resolutions using AMR. The solid lines with square signs
      are HLL-PLM. The dashed line indicates first order of global
      convergence.}\label{fig:l1_transverse_e}
\end{figure}

The results with a uniform grid and two AMR runs are shown in
Fig. \ref{fig:transverse_e}. The hydro solver used for this run is
HLL-PLM.  The CFL number used is $0.5$. The $L_1$ errors are shown in
Fig. \ref{fig:l1_transverse_e}.

Table 1 shows the equivalent resolution and the actual number of cells
used for this and the previous tests. It can be seen for the highest
resolution calculation our code uses about four hundred times less
grid cells than the corresponding uniform grid calculation. Thus AMR
allows us to achieve very high resolution while significantly reducing the computational
cost.

\begin{figure}
  \epsscale{1.3}
  \plotone{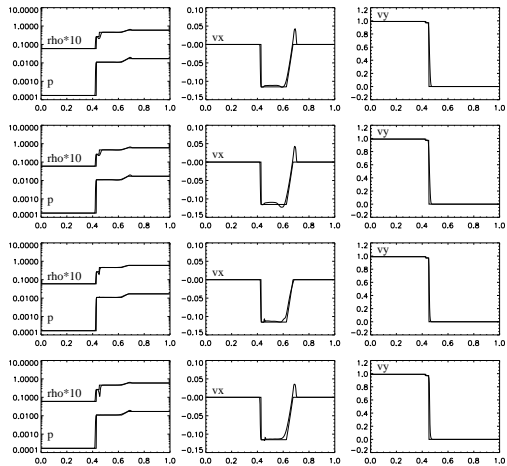}
    \caption{Jet-cocoon interaction problem at $t=0.4$ for (from top
      to bottom): HLL-PLM, LLF-PLM, HLLC-PLM and F-PLM. Thick lines
      are numerical solution while thin lines are analytical
      solutions. The grid resolution is uniform $N=400$. The CFL
      number is $0.5$.}\label{fig:jetcocoon400}
\end{figure}

\subsubsection{Jet-Cocoon interaction}
\label{sec:jet-cocoon}

For this test we set up a one-dimensional Riemann problem that mimics
the interaction of jet with an overpressured cocoon. The initial left
and right states for this problem are $p_L = 0.00017, \rho_L = 0.01,
v_{xL}=0, v_{yL} = 0.99$ and $p_R = 0.017, \rho_R = 0.1, v_{xR} = 0,
v_{yR} = 0$. The gas is assumed to be ideal with $\Gamma=5/3$. Those
values mimic the conditions of the jet-cocoon boundary in model C2 of
\citet{marti97}. The result using four different Riemann solver with
PLM on uniform grid are shown in Fig. \ref{fig:jetcocoon400}. It can
be seen that solutions use HLL, LLF and direct flux reconstruction
have large positive fluctuations in the normal velocity at the
rarefaction wave.  It is interesting to note that only HLLC does not
suffer from this shortcoming which is probably due to the ability of
HLLC to resolve the contact discontinuity compared to the other Riemann
solvers in the code. If those fluctuations also happen in higher
dimensional jet simulation, then one would expect that the normal
velocity fluctuation seen in this test would lead to an artificially
extended cocoon.



\begin{figure}
  \epsscale{1.3}
  \plotone{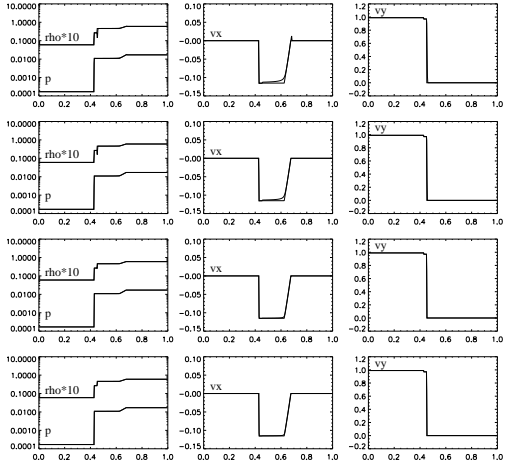}
    \caption{Jet-cocoon interaction problem at $t=0.4$ for HLL-PLM and
      HLLC-PLM. The top two are HLL-PLM and HLLC-PLM with 4 refinement
      levels and a refinement factor of 2 (equivalent resolution
      6400). The bottom two are HLL-PLM and HLLC-PLM with 4 refinement
      levels and a refinement factor of 4 (equivalent resolution
      102400). Thick lines are numerical solutions while thin lines are
      analytical solutions. The CFL number is
      $0.5$.}\label{fig:jet1d_amr}
\end{figure}
\begin{figure}
  \epsscale{1.2}
  \plotone{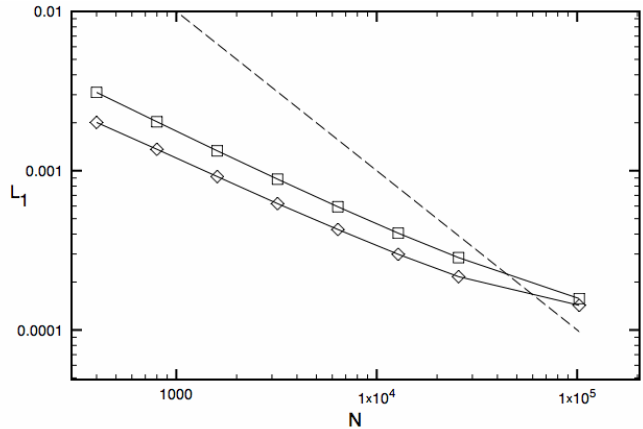}
    \caption{$L_1$ error for the jet-cocoon interaction problem for
      eight different grid resolutions using AMR. The square signs are
      for HLL-PLM while the diamonds are HLLC-PLM. The dashed line
      indicates first order of global
      convergence.}\label{fig:l1_jet1d}
\end{figure}

In Fig. \ref{fig:jet1d_amr} the result of using HLL-PLM and HLLC-PLM
with AMR is shown. It can been seen that the fluctuation in the HLL
scheme becomes smaller with higher resolution. Fig. \ref{fig:l1_jet1d}
shows the $L_1$ error for those two schemes with different levels of
refinement.

\subsection{Two Dimensional Test: Shock Tube}

To test the code in higher dimension, we first study the
two-dimensional shock tube problem suggested by \citet{del zanna02}
and latter also used by various groups (see e.g. Zhang \& MacFadyen
2006, Mignone \& Bodo 2005a, Lucas-Serrano et al. 2004). This test is
done in a two-dimensional Cartesian box divided into four
equal-area-constant states:
\begin{eqnarray}
(\rho, v_x, v_y, p)^{NE} = (0.1, 0, 0, 0.01), \\ \nonumber
(\rho, v_x, v_y, p)^{NW} = (0.1, 0.99, 0, 1), \\ \nonumber
(\rho, v_x, v_y, p)^{SW} = (0.5, 0, 0, 1), \\ \nonumber
(\rho, v_x, v_y, p)^{SE} = (0.1, 0, 0.99, 1), \nonumber
\end{eqnarray}
where NE means northeast corner and so on. The grid is uniform
$400\times400$. The gas is assumed to be ideal with an adiabatic index
$\Gamma=5/3$. We use outflow boundary conditions in all four
directions and the CFL number is $0.5$.

\begin{figure}
  \epsscale{1.3}
  \plotone{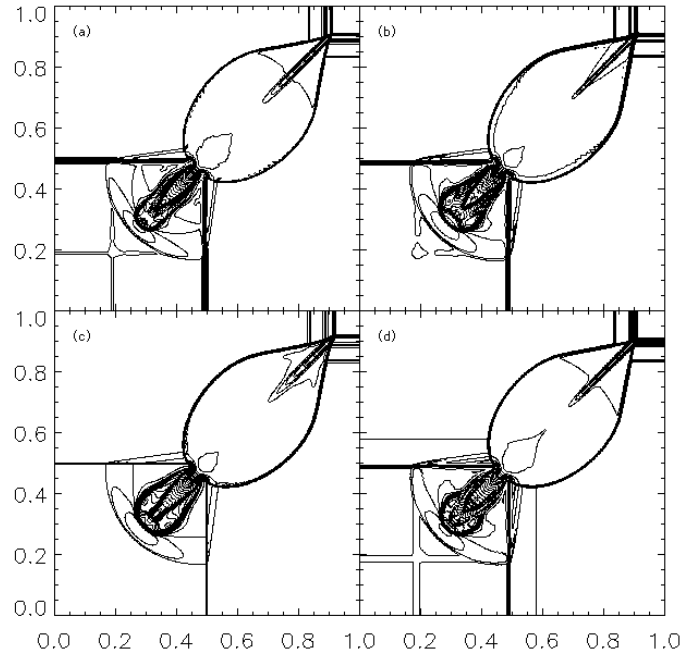}
    \caption{Shock tube problem at $t=0.4$ for (a) HLL-PLM, (b)
      HLL-PPM, (c) HLLC-PLM and (d) F-WENO5. Thirty equally spaced
      contours of the logarithm of density are plotted. The CFL number
      is $0.3$.}\label{fig:shocktube}
\end{figure}

The results are shown in Fig. \ref{fig:shocktube} for four
schemes. This problem does not have analytical solutions to compare
with, but comparing our result with other groups' result shows good
agreement. The cross in the lower left corners of (a) HLL-PLM and (d)
F-WENO is a numerical artifact due to the inability to maintain a
contact discontinuity perfectly, which are absent in the results using
the HLLC solver (c). This agreed with the result of \citet{mignone05a}
that the HLLC solver behaves better in this problem than other Riemann
solvers because of its ability of resolve contact discontinuities.

\begin{figure}
  \epsscale{1.4}
 \plotone{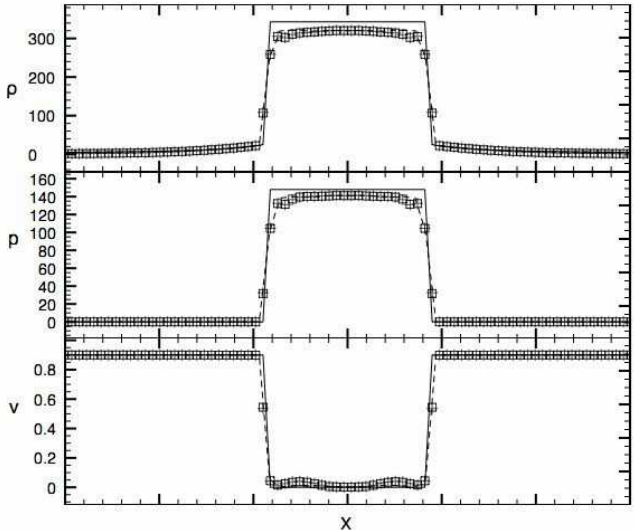}
    \caption{One-dimensional cut of the relativistic spherical shock
      reflection problem at $t=0.4$ for density (top), pressure
      (middle) and radial velocity (bottom). The solid line is the
      analytical solution. Square signs, plus signs and dashed lines
      are cut though x axis, y axis and diagonal direction,
      respectively. The resolution is a uniform grid with $128^3$
      zones.  LLF-PLM are used with a CFL number of $0.1$. It can be
      seen that the fluctuations in density and pressure at the sphere
      center present in many previous codes are absent in our
      results.}\label{fig:rssr_1d}
\end{figure}

\subsection{Three Dimensional Tests}

\subsubsection{Relativistic Spherical Shock Reflection}

We first study a test problem with uniform grid in Cartesian coordinate that has been used by
several groups to test the symmetric properties of a three-dimensional
SRHD code \citep{aloy99, mignone05b}. The initial setup of this test
consists of a cold spherical inflow with initially constant density
$\rho_0$ and constant velocity $v_1$ colliding at the box center. This
problem is run using three-dimensional Cartesian coordinates so it
allows one to evaluate the symmetry properties of the code
\citep{marti97, aloy99, mignone05b}.  When gas collides at the center,
a reflection shock forms. Behind the shock, the kinetic energy
will be converted completely into internal energy. Thus the downstream
velocity $v_2$ is zero and the specific internal energy is given by
the upstream specific kinetic energy
\begin{equation}
\epsilon_2 = W_1-1.
\end{equation}
Using the shock jump condition, the compression ratio $\rho_2/\rho_1$
and the shock velocity can be found to be
\begin{eqnarray}
{\rho_2\over \rho_1}={\Gamma+1\over\Gamma-1}+{\Gamma\over
  \Gamma-1}\epsilon_2 \\ V_s = \frac{(\Gamma-1)W_1|v_1|}{W_1+1}.
\end{eqnarray}

In the unshocked region ($r>V_st$), the gas flow will develop a
self-similar density distribution,
\begin{equation}
\rho_1=\left(1+{v_1t\over r}\right)^2\rho_0 . \label{rssr1}
\end{equation}

\begin{figure}
  \plotone{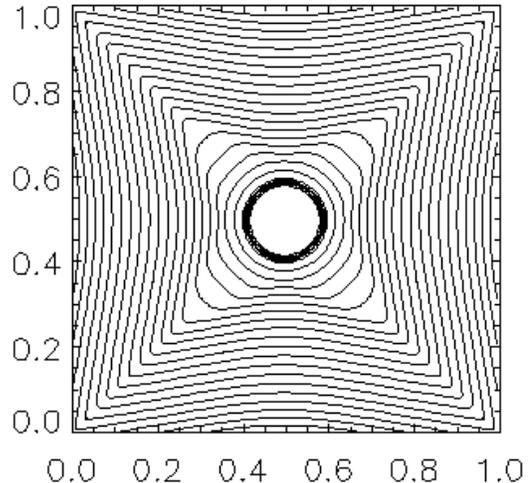}
    \caption{Density contour through $z=0.5$ of of the relativistic
      spherical shock reflection problem at $t=0.4$. It can be seen
      that the symmetry of the initial condition is preserved rather
      well up to this time. After this time, the unphysical boundary
      condition will begin to affect the subsequent
      evolution. }\label{fig:rssr_2d}
\end{figure}

The initial state are $p_1=7\times 10^{-6}, \rho_0= 1, v_1=-0.9$. We
chose a small value for pressure because a grid-based code cannot
handle zero pressure. A CFL number $0.1$ is used for this problem, as
other groups \citep{aloy99}. We chose to use LLF-PLM for this problem
because this turns out to be the most stable solver for this
problem. Fig. \ref{fig:rssr_1d} shows the one-dimensional cut though
axis and diagonal direction and Fig. \ref{fig:rssr_2d} shows a contour
through $z=0.5$ plane, both at $t=0.4$. It can be seen from those
plots that our code keeps the original spherical symmetry quite
well. Since in a Cartesian box the simple outflow boundary condition
is inconsistent with the initial spherical inflow setup, we evolve
this problem only to $t=0.4$, at which point all the mass in the
original box has just entered the shocked region (see
Fig. \ref{fig:rssr_1d}). After that time, the evolution would be
affected by the unphysical boundary condition.

\subsubsection{Relativistic Blast Wave I}

In this test, we study a spherical blast wave in three-dimensional
Cartesian coordinates. There is no analytical solution for this
problem.  Thus for the sake of comparison, we set up the same problem
as other groups \citep{del zanna02, zhang06}. The center of the blast
wave source is located at the corner $(0, 0, 0)$ of the box. The
initial conditions are
\begin{eqnarray}
(\rho, v_r, p) = (1, 0, 1000) \ \ {\rm if }  \ \ r\le0.4, \\ \nonumber
(\rho, v_r, p) = (1, 0, 1) \ \ {\rm if } \ \ r>0.4, \nonumber
\end{eqnarray}
where $r$ is the distance to the center $(0,0,0)$.

An ideal gas with an adiabatic index of $\Gamma=5/3$ is assumed. The
left boundaries at x, y, z directions are reflecting while others are
outflow. We use a top grid of $128^3$ zones with two more levels of
refinement and a refinement factor of 2 for this calculation
(equivalent resolution $512^3$). F-PLM is used for the result shown
and the CFL number is $0.1$.

\begin{figure}
  \plotone{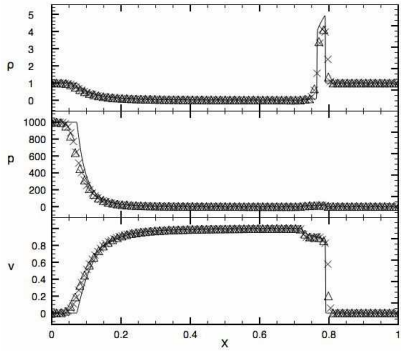}
    \caption{One-dimensional cut of the three-dimensional blast wave
      problem I at $t=0.4$ for density (top), pressure (middle) and
      velocity (bottom). The triangle signs and cross signs are cuts
      along the x axis and diagonal direction, respectively. The solid
      line is a one-dimensional simulation run using spherical
      coordinates with $4000$ uniform zones. F-PLM is used for this
      calculation. A top grid of $128^3$ zones with two levels of
      refinement and a refinement factor of 2 is used. The CFL number is
      $0.1$.}\label{fig:blast3dcorner_1}
\end{figure}

The results are given in Fig. \ref{fig:blast3dcorner_1} which shows
the cut along x-axis and diagonal direction.  For comparison, we run a
high resolution one-dimensional simulation using spherical
coordinates.  The three-dimensional run in Cartesian coordinates
agrees with the one-dimensional high resolution run. Furthermore, it
can be seen that the spherical symmetry of the initial condition is
preserved rather well in the three-dimensional Cartesian case.

\subsubsection{Relativistic Blast Wave II}

Finally we study another blast wave problem for which the center of
the blast wave source is located at the box center. This problem also
does not have analytical solution but it has been studied by
\citep{hughes02} so our result can be compared to theirs. The initial
conditions are
\begin{eqnarray}
(\rho, v_r, p) = (1, 0, 10^4)  \ \ {\rm if} \ \ r \le 0.05, \\ \nonumber
(\rho, v_r, p) = (0.1, 0, 10) \ \ {\rm if} \ \ r > 0.05. \nonumber
\end{eqnarray}
An ideal gas EOS with an adiabatic index of $\Gamma=4/3$ is used. We stop
the run at $t=0.12$, roughly the same ending time of \citet{hughes02}. A
top grid of $64^3$ zones with four levels of refinement and a refinement factor
of two is used (equivalent resolution $1024^3$). Thus our resolution is
roughly $1.5$ times of \citet{hughes02}. We used HLL-PLM and a CFL
number of $0.5$ for this calculation.

\begin{figure}
  \plotone{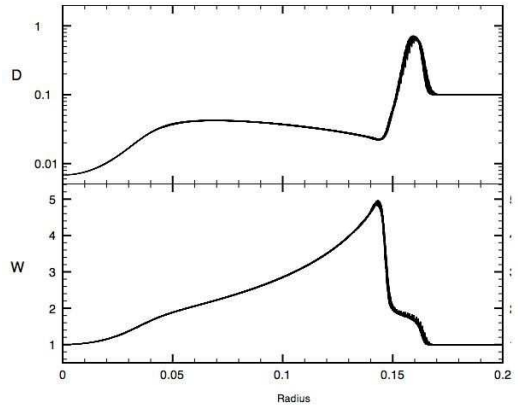}
    \caption{Three-dimensional blast wave problem II at $t=0.12$.  All
      points in a slice at $y=0.5$ are plotted as a function of the
      distance to the center $(0.5,0.5,0.5)$. The upper panel shows
      the laboratory density $D$ and the lower panel shows the Lorentz
      factor $W$. HLL-PLM is used for this calculation. A top grid of
      $64^3$ zones with four levels of refinement and a refinement
      factor of 2 is used. The CFL number is
      $0.5$.}\label{fig:blast3d_1}
\end{figure}

Fig.~\ref{fig:blast3d_1} plots the numerical solution for all cells
centered on the highest level in the two dimensional slice at $y=0.5$
at $t = 0.12$ as a function of radius from the center
(0.5,0.5,0.5).  The position and amplitude of the high density shell
agrees with the calculation of \citet{hughes02}. And it can be seen
that the spherical symmetry is preserved rather well in our code.


\section{Astrophysical Application I Relativistic Supersonic Jet Propagation}
\label{sec:jet3d}

\begin{figure*}[!t]
\resizebox{\textwidth}{!}{\rotatebox{0}{\plotone{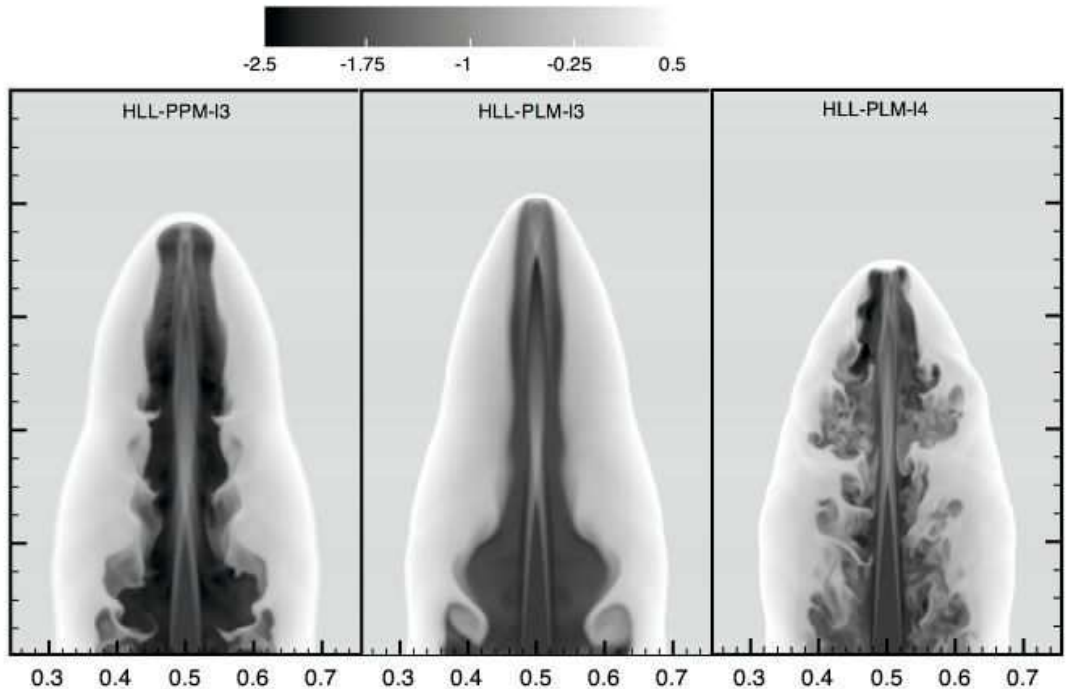}}}
\caption{Density slice through $y=0.5$ for three dimensional
  relativistic jet at t=$60 r_b/c$ for HLL-PPM with three levels of
  refinement (left), HLL-PLM with three levels of refinement (middle)
  and HLL-PLM with four levels of refinement (right). The top grid
  resolution is $64^3$ zones and a refinement factor of 2 is
  used. Thus the left and middle panels have an equivalent resolution
  of $512^3$ zones while the right one has $1024^3$ zones. The CFL
  number is $0.4$.}\label{fig:jet}
\end{figure*}

Having validated our code using various test problems, we can apply
it to astrophysical problems. We will study two typical astrophysical relativistic
 flow problems in this work, AGN jets and GRB jets. Both topics have been studied 
 extensively with 2d simulations before, but very few 3d calculations have been done 
 in both cases. Consequently we will focus on 3d simulations here.

In this section, we study a relativistic supersonic jet in three
dimensions. We set up the problem using the same parameters as model
C2 of \citep{marti97}.  This model has also been studied in two
dimensions by \citet{zhang06} and in three dimensions by
\citet{aloy99}. The jet parameters are $\rho_b=0.01$, $v_b=0.99$ and
$r_b=0.02$. The jet has a classical Mach number $M_b=v_b/c_s=6$ so the
pressure is $p_b=0.000170305$. The parameters for the medium are
$\rho_m=1.0$, $v_m=0$ and $p_m=p_b$. The EOS is assumed to be ideal
with $\Gamma=5/3$. The jet is injected from the low-z boundary
centered at $(0.5, 0.5, 0)$ with radius $r_b$. Outflow boundary
conditions are used at other part of the boundary.

Figure \ref{fig:jet} shows the result at $t=60r_b/c$ for the three
runs: HLL-PPM with three levels of refinement (HLL-PPM-l3), HLL-PLM
with three levels of refinement (HLL-PLM-l3) and HLL-PLM with four
levels of refinement (HLL-PLM-l4). The top grid resolution is $64^3$
zones.  Thus the first two runs have an equivalent resolution of
$512^3$ zones while the last one has $1024^3$ zones. For the first
two, turbulence in cocoon is not fully developed so the cocoon is
still symmetric even in 3d. The HLL-PPM-l3 run has slightly more
turbulent cocoon due to the higher spatial reconstruction order of
PPM.  Thus the HLL-PPM-l3 jet propagates slightly slower than the
HLL-PLM-l3 jet. On the other hand, for the HLL-PLM-l4 jet, the
resolution is $20$ cells per beam radius, comparable to the resolution
used in the two-dimensional study by \citet{marti97}. The cocoon
turbulence is much more developed in this case, as in the 2d
case \citep{marti97}. Consequently, the HLL-PLM-l4 jet propagates
slower than the two lower resolution ones. Furthermore, the HLL-PLM-l4
case does not show axisymmetry because instability quickly develops in
the lateral motion and consequently lateral motion also becomes
turbulent.

Since we found the jet-cocoon structure differs significantly at higher resolution
run and consequently the jet propagation speed decreases, we conclude that
even at $1024^3$ effective resolution of our three dimensional jet
simulations the correct solution remains elusive. Moreover, different
solvers give disparate answers.

\section{Astrophysical Application II. Three Dimensional Simulation of Collapsar Jets}

\begin{figure}
 \plotone{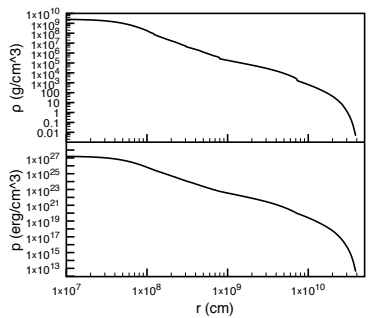}
    \caption{Density profile and pressure profile of the HE16A progenitor model adopted in our GRB 
    jet calculation.} \label{star}
\end{figure}

\begin{figure*}[!t]
\resizebox{\textwidth}{!}{\rotatebox{0}{\plotone{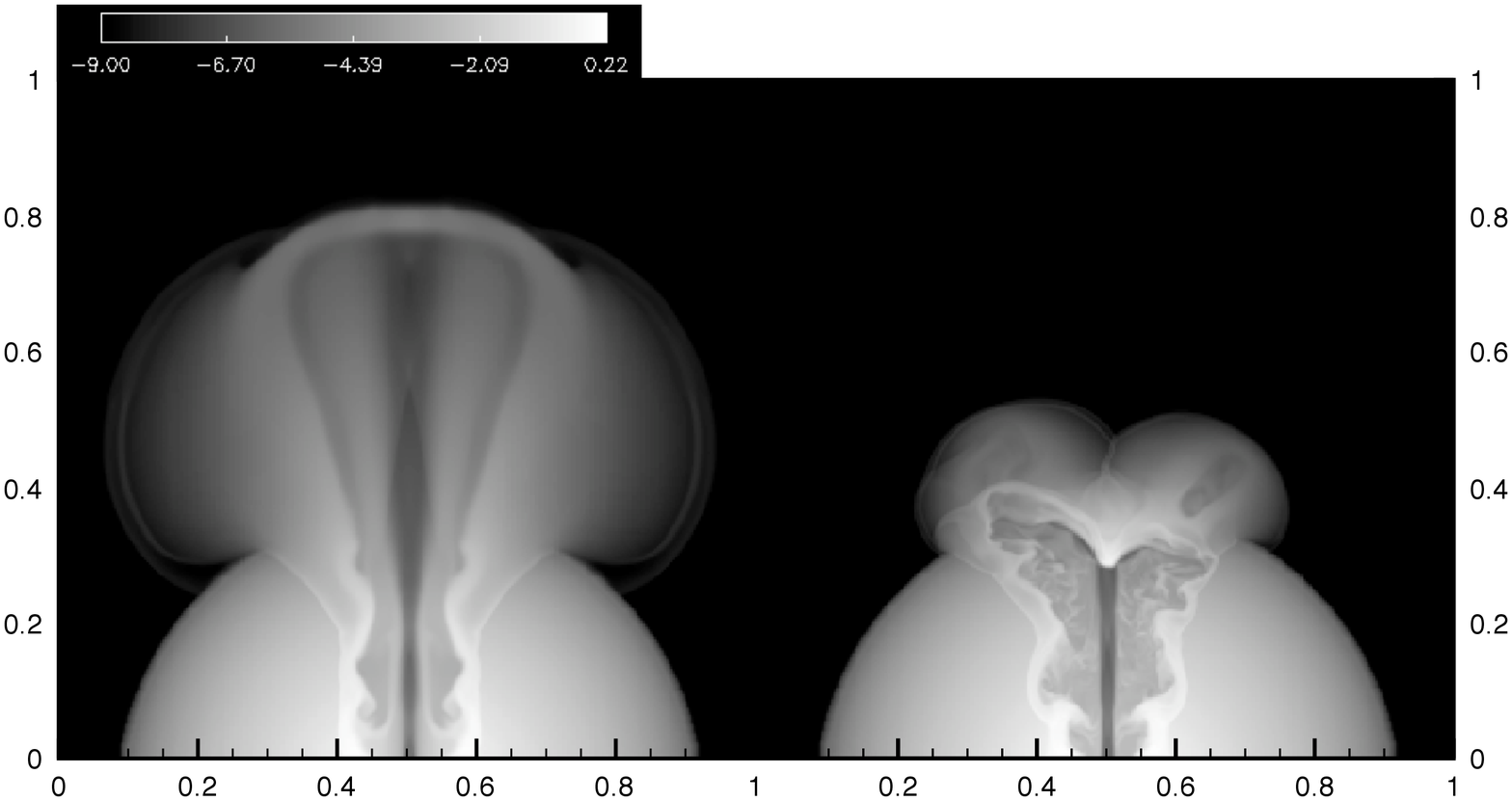}}}
\caption{Density slice through $y=0.5$ for three dimensional
  relativistic GRB jet at t=$10.99$ s for HLL-PLM with four levels of
  refinement (left) and five levels of refinement (right). The top grid
  resolution is $128^3$ zones and a refinement factor of 2 is
  used. Thus the left and middle panels have an equivalent resolution
  of $2048^3$ zones while the right one has $4096^3$ zones. The CFL
  number is $0.4$.}\label{grbjet}
\end{figure*}

The idea that ``long-soft" gamma-ray bursts are associated with the deaths of massive stars
has been supported by various observations recently (see e.g. Woosley \& Bloom (2006) for a 
recent review). So it is of great interest to calculate how a relativistic jet propagate through and break
out of a massive star. There has been extensive calculations in 2d recently \citep{zhang03, zhang04, 
nagataki06, morsony06}. But very few 3d calculation have been reported in the literature so far. Since there are
significant turbulent motion and mixing in the jet-cocoon 
system in 2d calculation, it is important to model this process in 3d.

We use the projenitor HE16A of \citet{woosley06b}, which is a stripped-down helium core with initial mass $16$ M$_\odot$. Fig.~\ref{star} shows the density and pressure profiles for the progenitor model. The radius of the star at onset of collapse is $3.86\times 10^{10}$ cm. In our simulation, the mass inside $5\times10^9$ cm is removed and replaced by a point mass of $9.3$ M$_\odot$. The jet has power $\dot E=3\times 10^{50}$ erg s$^{-1}$, initial Lorentz factor $\Gamma_0=5$, the ratio of its total energy (excluding rest mass energy) to its kinetic energy $f_0=40$. This corresponds to a a jet with initial density $5.937\times10^{-3}$ g cm$^{-3}$ and pressure $4.156\times10^{19}$ erg cm$^{-3}$. The jet is injected parallel to the $z$ axis with an initial radius $8.73\times10^8$ cm. Those model parameters are similar to previous 2d calculations \citep{zhang04, nagataki06}. 

We use a simulation box of $3.2\times10^{11}$ cm in order to follow the propagation of the jet after break out. We use HLL-PLM as this should be the most stable and reliable scheme to carry out resolution study. The top grid resolution is $128^3$. We run the simulation using both 4 and 5 refinement levels, which correspond to resolution of 5.6 and 11 cells per jet beam radius, respectively. In order to always have high resolution for the jet material, we designed a color field refinement strategy in addition to the standard refinement criterion designed for discontinuities. More specifically, we use two color fields to keep track of the injected jet material and the stellar material. Those two color fields give us a fraction of jet material at every cell. Then whenever a cell contains more than 0.1 percent of jet material, we flag that cell for refinement. This ensures that we also have high resolution when mixing between jet material and star material happens.

The results for those two runs are show in Fig. \ref{grbjet}. It can been seen that the high resolution run gives qualitatively different jet dynamics. While in the low resolution run the jet break out of the star successfully, in the high resolution run the jet head bifurcate at the stellar edge. Interestingly, this behavior has also been seen in 2d calculation \citep{morsony06}. But since we did not get convergent behavior so far, we cannot conclude at this stage whether this behavior is physical or purely numerical. However, it is safe to conclude that much higher resolution will be needed to model the jet break out in 3d.

\section{Conclusions and Discussions}

In this paper, we have described a new code that solves the special
relativistic hydrodynamics equations with both spatially and
temporally adaptive mesh refinement. It includes direct flux
reconstruction and four approximate Riemann solvers including HLLC,
HLL, LLF and modified Marquina flux formula. It contains several
reconstruction routines: PLM, PPM, third order CENO, and
third and fifth order WENO schemes. A modular code structure
makes it easy to include more physics modules and new algorithms.

From our test problems and two astrophysical applications, it is clear that relativistic flow problems
are more difficult than the Newtonian case. One key reason is that
in the presence of utrarelativistic speed, nonlinear structures such
as shocked shells are typically much thinner and thus requires the use
of very high spatial resolution. SRHD problems also become difficult to
solve accurately when significant transverse velocities are present in
the problem as we have shown using several one dimensional
problems. One reason for this difficulty is that in SRHD velocity
components are coupled nonlinearly via the Lorentz factor.  In
studying astrophysical jet problems, we have demonstrated the need of
both high resolution achievable only through AMR and careful choice of
hydrodynamic algorithms. In addition to validate our AMR code, the most important implications of 
the calculations we have done is that \emph{in relativisitic flow simulations, resolution studies are crucial}.

%



\acknowledgments

We thank Greg Bryan and Michael Norman for sharing {\sl enzo} with the
astrophysical community without which this study could have not been
carried out. We would also like to thank Miguel Aloy for very helpful
comment on the draft, Roger Blandford and Lukasz Stawarz for helpful
discussions. Furthermore, we also thank Ralf K\"aehler for help with
optimizing the code and SLAC's computing servers for maintaining an
SGI Altix super computer on which the reported calculations were
carried out. This work was partially supported by NSF CAREER award
AST-0239709 and 
Grant No. PHY05-51164 from the National Science Foundation. P. W. acknowledges
support by the Stanford Graduate Fellowship and KITP Graduate Fellowship. W. Z. has been supported
by NASA through Chandra Postdoctoral Fellowship PF4-50036 awarded by
the {\sl Chandra X-Ray Observatory} Center, and the DOE Program for
Scientific Discovery through Advanced Computing (SciDAC).



\end{document}